%Paper: hep-th/9305185
%From: jhs@theory3.caltech.edu (John Schwarz)
%Date: Mon, 31 May 93 14:02:43 PDT

\input phyzzx

\let\refmark=\NPrefmark  
\def\define#1#2\par{\def#1{\Ref#1{#2}\edef#1{\noexpand\refmark{#1}}}}
\def\con#1#2\noc{\let\?=\Ref\let\<=\refmark\let\Ref=\REFS
         \let\refmark=\undefined#1\let\Ref=\REFSCON#2
         \let\Ref=\?\let\refmark=\<\refsend}

\define\RTOWN
J. Azcarraga, J. Gauntlett, J. Izquierdo and
P. Townsend, Phys. Rev. Lett. {\bf 63} (1989) 2443.

\define\RNEPO
R. Nepomechie, Phys. Rev. {\bf D31} (1984) 1921;
C. Teitelboim, Phys. Lett. {\bf B167} (1986) 69.

\define\RRECENT
R. Kallosh and T. Ortin, preprint SU-ITP-93-3 (hep-th/9302109);
P. Bin\'etruy, preprint NSF-ITP-93-60; M. Duff and R. Khuri, preprint
CTP/TAMU-17/93 (hep-th/9305142).

\define\RFSSS
A. Font, L. Ib\'a\~nez, D. Lust and F. Quevedo, Phys. Lett. {\bf B249} (1990)
35;
S.J. Rey, Phys. Rev. {\bf D43} (1991) 526;
A. Shapere, S. Trivedi and F. Wilczek, Mod. Phys. Lett. {\bf A6}
(1991) 2677.

\define\RCECO
S. Cecotti, S. Ferrara and L. Girardello, Nucl. Phys. {\bf B308} (1988)
436;
M. Duff, Nucl. Phys. {\bf B335} (1990) 610;
J. Molera and B. Ovrut, Phys. Rev. {\bf D40} (1989) 1146;
T. Kugo and B. Zwiebach, Prog. Theor. Phys. {\bf 87} (1992) 801.

\define\RNARAIN
K. Narain, Phys. Lett. {\bf B169} (1986) 41;
K. Narain, H. Sarmadi and E. Witten, Nucl. Phys. {\bf B279} (1987) 369.

\define\RSLTZ
A. Sen, preprint TIFR-TH-93-03 (hep-th/9302038),
to appear in Int. J. Mod. Phys. A.

\define\RSCHSEN
J. Schwarz and A. Sen, preprint NSF-ITP-93-46 (hep-th/9304154).

\define\ROLIVE
C. Montonen and D. Olive, Phys. Lett. {\bf B72} (1977) 117;
P. Goddard, J. Nuyts and D. Olive, Nucl. Phys. {\bf B125} (1977) 1;
H. Osborn, Phys. Lett. {\bf B83} (1979) 321.

\define\RHARLIU
J. Harvey and J. Liu, Phys. Lett. {\bf B268} (1991) 40; R. Kallosh, A.
Linde, T. Ortin, A. Peet and A. Van Proeyen, Phys. Rev. {\bf D46} (1992)
5278; T. Ortin, preprint SU-ITP-92-24 (hep-th/9208078).

\define\RSTROM
A. Strominger, Nucl. Phys. {\bf B343} (1990) 167; C. Callan, J. Harvey and
A. Strominger, Nucl. Phys. {\bf B359} (1991) 611; {\bf B367} (1991) 60;
preprint EFI-91-66 (hep-th/9112030).

\define\RDUFF
M. Duff, Class. Quantum Grav. {\bf 5} (1988) 189;
M. Duff and J. Lu, Nucl. Phys. {\bf B354} (1991) 129, 141;
{\bf B357} (1991) 354;
Phys. Rev. Lett. {\bf 66} (1991) 1402;
Class. Quantum Grav. {\bf 9} (1991) 1;
M. Duff, R. Khuri and J. Lu, Nucl. Phys. {\bf B377} (1992) 281;
J. Dixon, M. Duff and J. Plefka, Phys. Rev. Lett. {\bf 69} (1992) 3009.

\define\RMAHSCH
J. Maharana and J. Schwarz, Nucl. Phys. {\bf B390} (1993) 3.

\define\RSCHWARZ
J. Schwarz, preprint CALT-68-1815 (hep-th/9209125).

\define\RDUALITY
A. Sen, preprint TIFR-TH-92-41 (hep-th/9207053) (to appear in Nucl.
Phys. B).

\define\RDIRAC
P. Dirac, Proc. R. Soc. {\bf A133} (1931) 60; J. Schwinger, Phys. Rev.
{\bf 144} (1966) 1087; {\bf 173} (1968) 1536; D. Zwanziger, Phys. Rev.
{\bf 176} (1968) 1480, 1489.

\define\RWITTEN
E. Witten, Phys. Lett. {\bf 86B} (1979) 283.

\define\RDYONO
A. Sen, Phys. Lett. {\bf B303} (1993) 22.

\define\RDYONT
A. Sen, preprint NSF-ITP-93-29 (hep-th/9303057).

\define\RTSEYTLIN
A. Tseytlin, Phys. Lett. {\bf B242} (1990) 163; Nucl. Phys. {\bf B350}
(1991) 395.

\define\RBOGOM
D. Olive and E. Witten, Phys. Lett. {\bf 78B} (1978) 97;
G. Gibbons and C. Hull, Phys. Lett. {\bf B109} (1982) 190.

\overfullrule=0pt
\hsize=16.5cm
\vsize=23.0cm
\hoffset=0.1cm

\def\tF{\tilde F}
\def\tj{\tilde j}

\def\ca{{\cal A}}

\def\hG{\hat G}
\def\hB{\hat B}
\def\cm{{\cal M}}

\def\cl{{\cal L}}
\def\p{\partial}

\def\hcl{\hat\cl}

\def\tJ{\tilde J}

{}~\vbox{\hbox{NSF-ITP-93-64}\hbox{CALT-68-1866}
\hbox{TIFR-TH-93-24}\hbox{hep-th/9305185}}\break

\title{DUALITY SYMMETRIES OF 4D HETEROTIC STRINGS}

\author{John H. Schwarz\foot{Supported in part by the U.S. Dept. of
Energy under Grant No. DE-FG03-92-ER40701.}
\foot{JHS@THEORY3.CALTECH.EDU}}

\address{Institute for Theoretical Physics, University of California,
Santa Barbara, CA 93106}

\address{California Insitute of Technology,
Pasadena, CA 91125\foot{Permanent address.}}

\andauthor{Ashoke Sen\foot{SEN@TIFRVAX.BITNET}\foot{Supported
in part by National Science Foundation under grant No. PHY89-04035.}}

\address{Institute for Theoretical Physics, University of California,
Santa Barbara, CA 93106}

\address{Tata Institute of Fundamental Research, Homi Bhabha Road,
Bombay 400005, India${}^{\ddagger}$}

\bigskip

\abstract

Target space duality (T duality), which interchanges Kaluza--Klein
and winding-mode excitations of the compactified heterotic string,
is realized as a symmetry of a world-sheet action. Axion-dilaton
duality (S duality), a conjectured nonperturbative SL(2,Z) symmetry
of the same theory, plays an analogous role for five-branes. We
describe a soliton spectrum possessing both duality symmetries and argue
that the theory has an infinite number of dual string descriptions.

\vfill
\endpage

\noindent{\bf 1. Introduction}

One of the major gaps in our understanding of string theory is the lack
of a fundamental formulation of the
nonperturbative theory.  Many efforts have been made to gain insight
into nonperturbative aspects of string theory in recent years.
These include studies of
matrix models, construction of soliton solutions such as black holes and
magnetic monopoles, studies of string field theory, and much more.
Another recent focus, which will be pursued here, is a
proposed nonperturbative SL(2,Z) symmetry of the heterotic string theory
compactified to four dimensions\con\RFSSS\RDUALITY\RSCHWARZ\noc.
While such a symmetry is not yet
definitively established, the evidence for it is certainly
mounting\con\RDYONO\RDYONT\RSCHSEN\noc. To
be concise, let us refer to this symmetry as {\it S duality}.

If present at all, S duality is necessarily nonperturbative, since it
transforms the four-dimensional dilaton field, whose value determines the
string loop expansion parameter (Newton's constant), nonlinearly.
Despite this fact, it has many remarkable similarities with target-space
duality (called {\it T duality}), which is also an infinite discrete group.
(This group generalizes the well-known $R \rightarrow 1/R$ symmetry.)
In the case of  toroidal compactification of the heterotic
string, in the manner originally proposed by Narain\RNARAIN,
$G_T$ = O(6,22;Z).
In general, the group $G_T$ depends on the particular compactification
chosen.  Other examples that have been studied include certain orbifolds
and Calabi--Yau spaces.  Unlike $G_T$, the S duality group SL(2,Z) seems
to be
``universal'' in the sense that it does not depend on the
compactification chosen, at least if the choice preserves some
supersymmetry in four dimensions.  Of course, in the case of toroidal
compactification (the only case we will consider explicitly), there is
N = 4 supersymmetry in four dimensions.

Heuristically, one can describe the toroidally compactified heterotic
string theory  by an effective
four-dimensional action, containing fields associated with massless
quanta only. Effects due to finite string size and string loops
are then represented as a double series expansion
in the string scale $\alpha'$ and Newton's constant.
Of course, these series do not converge, and there are
important nonperturbative phenomena associated with both expansions.
The leading term (in both senses), which is a classical N = 4, D = 4 field
theory, has both dualities --- O(6,22) and SL(2,R), but as usually
formulated, there is an apparent asymmetry between them.
Namely, O(6,22) is a manifest symmetry
of the action, whereas SL(2,R) is a symmetry of the equations of
motion only. However, in a recent paper\RSCHSEN\
we showed that it is possible to
recast the theory, by introducing suitable auxiliary fields, so that
both symmetries are realized simultaneously in the action in essentially
the same way.  In certain cases, the price for doing this is that the
action no longer has manifest general coordinate invariance, though this
symmetry is still present.  The way this works is that the formulas for
general coordinate transformations of vector fields are modified from
the usual ones by terms that vanish when the equations of motion are satisfied.

An analogous mathematical problem arises in understanding the T duality
group O(6,22) in the 2D world-sheet theory, which underlies the
$\alpha'$ expansion.  Namely, in the usual formulation O(6,22) is a
symmetry of the world-sheet field equations only, not the world-sheet
action.  In section 2, methods analogous to those
employed for the 4D problem are used to find a new form of the
world-sheet action possessing O(6,22) symmetry. (This generalizes
previous work by Tseytlin\RTSEYTLIN, which contained many of the essential
ideas.) The boundary condition on the world-sheet fields break this
O(6,22) symmetry to O(6,22;Z).
Thus, if there are no anomalies, the toroidally compactified heterotic
string theory
should have this symmetry order-by-order in Newton's constant, provided
that at each order the full nonperturbative $\alpha'$ dependence is
taken into account.  It seems plausible
that the corresponding statement can be
made for the S duality symmetry SL(2,Z) when the role of the $\alpha'$
and Newton's constant expansions are interchanged, i.e., S duality should be
true order-by-order in $\alpha'$ when the full nonperturbative Newton's
constant structure is included at each order.  This interchange in the
roles of $\alpha'$ and Newton's constant corresponds roughly
to what one gets by replacing the string theory by a dual
theory\RNEPO\ based on five-branes\RDUFF\RSTROM.
This is only heuristic, however, since there is no well-defined
quantum theory of five-branes as yet. %***
In any case, we propose to refer to this
expected symmetry between the roles of the two duality groups as
{\it duality of dualities}.

In section 3 a mass formula for string solitons as a function
of their electric and magnetic charges is described. By assuming
that a Bogomol'nyi bound is saturated (as is expected for an N = 4
theory), the spectrum of soliton masses is shown to
depend on the moduli in
just the right way to ensure O(6,22;Z) $\otimes$ SL(2,Z) symmetry.  The
spectrum of charges
corresponds to a 56-dimensional even self-dual lattice, whose
properties ensure that the
Dirac--Schwinger--Zwanziger--Witten\RDIRAC\RWITTEN\
(DSZW) quantization
requirements are automatically satisfied.  The states containing
electric charges only are present in the perturbative spectrum,
whereas all the others containing at least one non-zero magnetic charge
must arise nonperturbatively.
As a result, the perturbative spectrum is O(6,22;Z) invariant but not
SL(2,Z) invariant.
However, the spectrum of the perturbative five-branes,
compactified on a six-dimensional torus, contains states that carry both
magnetic and electric charges in such a way that the spectrum of
charges is symmetric under SL(2,Z) transformations, but not under
O(6,22;Z) transformations.

Since the string world-sheet theory does not have S duality, one
obtains a different world-sheet
theory by applying an SL(2,Z) transformation.
Section 4 explains that the transformed theories can be interpreted as
an infinite family of
isomorphic theories, any one of which provides an equally good starting
point for defining the full theory.  The essential difference between
different choices is which states in the spectrum
belong to the perturbative spectrum and which ones arise
nonperturbatively as solitons.  This SL(2,Z) duality of string theories
generalizes a Z$_2$ duality proposed for certain field theories by
Montonen and Olive\ROLIVE.
\medskip

\noindent{\bf 2. World-Sheet Action with Manifest O(6,22;Z) Symmetry}

When the heterotic string is compactified on a 28-torus that is conjugate to
an even self-dual lattice of signature (6,22), one obtains a consistent
four-dimensional theory.  The resulting 4D theory has N = 4
supersymmetry and contains the following massless bosons:  graviton
($g_{\mu\nu}$), 4D dilaton ($\Phi$),
antisymmetric tensor $(B_{\mu\nu})$ -- related by a duality transformation
to the axion ($\chi$), 28 abelian vector fields $(A_\mu^a)$ transforming as a
vector of the group O(6,22), and scalars (or moduli) described by a matrix
$M^{ab}$, which parametrizes the coset O(6,22)/O(6) $\times$ O(22).
The matrix $M$ is an arbitrary real symmetric $28
\times 28$ matrix belonging to the group O(6,22). The axion and dilaton
can be combined into a complex field $\lambda = \chi + ie^{-\Phi}
\equiv \lambda_1+i\lambda_2$,
which transforms under S duality
according to $\lambda \rightarrow (a\lambda + b)/(c\lambda
+ d)$, where $\left({a\ b\atop c\ d}\right) \in$ SL(2,Z).
For ``generic'' values of the moduli this is the complete massless
bosonic spectrum.  However, for special values corresponding to various
hypersurfaces in moduli space, there are additional massless states and
nonabelian gauge symmetries.  Certain parts of our analysis are not
easily generalized to include nonabelian gauge symmetries, so we
restrict the moduli to ``generic'' values.  The lifting of this
restriction is an important topic for future study.  To keep formulas
from becoming unwieldy, all fermions are dropped, though
their inclusion would not be an essential complication.

Let us now consider the string world-sheet theory in the presence of all
the bosonic fields listed above, each being a function of the
four-dimensional space-time coordinate $x^\mu$.  Thus, the world-sheet
theory contains as ``couplings'' exactly those fields that are included
in the low-energy effective field theory.  In the usual
formulation of the world-sheet theory, the moduli described by $M^{ab}$
appear in three distinct pieces corresponding to internal components of
the ten-dimensional metric, antisymmetric tensor, and vectors (of which
there are 16).  This action certainly does not have O(6,22) symmetry.
The equations of motion of the world-sheet theory can be recast in a
manifestly O(6,22) symmetric form,
however\RCECO\RTSEYTLIN\RMAHSCH.
For this purpose one introduces 28
world-sheet fields $y^a (\sigma,\tau)$ to parametrize the 28-torus
discussed earlier.  Since the geometric data reside in the moduli, each
$y^a$ can be regarded as an angular coordinate for a circle of unit
radius.  The invariant metric of the group O(6,22) is conveniently
taken to have the form
$$	L = \pmatrix{0 & I_6 & 0\cr
	I_6 & 0 & 0\cr
	0 & 0 & -I_{16}\cr}\,\, , \eqn\twoone$$
so that six eigenvalues are $+1$ and $22$ are $-1$.  Since $M^T LM=L$
and $M^T=M$, $M^{-1} = LML$.  In terms of these quantities,
it was shown in ref.\RMAHSCH\ that the world-sheet field equations can be
recast in the manifestly O(6,22) symmetric form\foot{The vectors
$A_\mu^{m+6}$ and $y^{m+6}$ ($1\le m\le 6$) are related to the vectors
$A^{(2)}_{m \mu}$ and $y^{m+6}$ of ref.\RMAHSCH\ by a minus sign.}
$$	D_0 y^a = -(ML)^a_{~b} D_1 y^b \eqn\twotwo$$
and
$$	\eqalign{g_{\mu\nu} \partial^\alpha \partial_\alpha x^\nu + &
\Gamma_{\mu\nu\rho} \partial^\alpha x^\nu \partial_\alpha x^\rho
	= - {1\over 2} D_1 y^a (L \partial_\mu M L)_{ab} D_1 y^b \cr &
	- \epsilon^{\alpha\beta} \partial_\alpha x^\nu F_{\mu\nu}^a
L_{ab} D_\beta y^b + {1\over 2} \epsilon^{\alpha\beta} H_{\mu\nu\rho}
\partial_\alpha x^\nu \partial_\beta x^\rho .\cr} \eqn\twothree$$
In these equations
$$	\eqalign{F_{\mu\nu}^a &= \partial_\mu A_\nu^a - \partial_\nu
A_\mu^a\cr
	H_{\mu\nu\rho} &= \partial_\mu B_{\nu\rho} + {1\over 2} A_\mu^a
L_{ab} F_{\nu\rho}^b + {\rm cyc. \ perms.}\cr
	D_\alpha y^a &= \partial_\alpha y^a + A_\mu^a
	\partial_\alpha x^\mu\cr}\eqn\twofour$$
	and $\Gamma_{\mu\nu\rho}$ is the usual Christoffel connection.
The $\int\Phi R^{(2)}d^2\sigma$ term has been dropped from the
world-sheet action, since it is higher order in $\alpha'$. %***
Note that
O(6,22) symmetry requires regarding the coordinates $y^a$ as a 28-vector.
Since they describe a product of 28 circles, it is clearly only possible
to rotate them with integer coefficients, so the group must be
restricted to O(6,22;Z).  Note also that $D_\alpha y^a$ is gauge
invariant provided that under a gauge transformation, $\delta A_\mu^a =
\partial_\mu \Lambda^a$, the internal coordinates transform as follows:
$\delta y^a = - \Lambda^a$.
Since the matrix $ML$ has 22 eigenvalues that are $-1$ and 6 that are
$+1$ the $y$ equation of motion \twotwo\ describes 22 left-moving bosons and 6
right-moving bosons.

Following Tseytlin\RTSEYTLIN\
(whose work we are generalizing here), it is
possible to find an action based on the world-sheet coordinates $x^\mu$
and $y^a$ that has manifest O(6,22) symmetry.  The Lagrangian (for
flat world-sheet metric) that gives the equations of motion \twotwo\ and
\twothree\ is
$$	\eqalign{{\cal L} =\ &{1\over 2} g_{\mu\nu} \eta^{\alpha\beta}
\partial_\alpha x^\mu \partial_\beta x^\nu
	- {1\over 2} D_0 y^a L_{ab} D_1 y^b - {1\over 2} D_1 y^a
(LML)_{ab}
D_1 y^b\cr
	&+ {1\over 2} \epsilon^{\alpha\beta} [B_{\mu\nu}
\partial_\alpha x^\mu \partial_\beta x^\nu - A_\mu^a \partial_\alpha
x^\mu L_{ab} D_\beta y^b] \,\, . \cr} \eqn\twofive$$
The $[U(1)]^{28}$ gauge invariance of this formula involves an interplay
between the last two terms, since $\delta B_{\mu\nu} =- {1\over 2}
F_{\mu\nu}^a L_{ab} \Lambda^b$.

To understand this theory better, it is important to exhibit the
coupling to a world-sheet metric $h_{\alpha\beta}$ that gives 2D Weyl
invariance and reparametrization invariance.  This is achieved by
replacing the first term (as usual) by
$$	{\cal L}'_1 = {1\over 2} \sqrt{-h}  h^{\alpha\beta} g_{\mu\nu}(x)
\partial_\alpha x^\mu \partial_\beta x^\nu\,\, , \eqn\twosix$$
and the third term by
$$	{\cal L}'_3 =- {1\over 2\sqrt{-h} h^{00}} D_1 y^a
(LML)_{ab}
D_1 y^b - {h^{01}\over 2 h^{00}} D_1 y^a L_{ab} D_1 y^b \,\, .
\eqn\twoseven$$
The other three terms are not modified.  The last two terms are
reparametrization and Weyl invariant as they stand.  The second term
(${\cal L}_2$) is not reparametrization
invariant, but it turns out the sum ${\cal L}_2 + {\cal L}'_3$ is.
The construction used here is a specialization to two dimensions of the
general method introduced in ref.\RSCHSEN.
Reparametrization invariance is achieved by modifying the usual rule
$\delta y^a = \xi^1 \partial_1 y^a + \xi^0 \partial_0 y^a$.
Specifically, the term $\partial_0 y^a$ should be replaced by the
expression that it equals as a result of the $y^a$ equation of motion.
All other transformations are the usual ones.

The action can now be varied with respect to $h^{\alpha\beta}$ to give
the  symmetric traceless energy--momentum tensor $T_{\alpha\beta}$.
The requirement that $T_{\alpha\beta}$ vanishes
gives the usual Virasoro conditions,  which are rather simple in the
$h_{\alpha\beta} = \eta_{\alpha\beta}$ gauge.  Alternatively, if one
wishes, the $h_{\alpha\beta}$ equations of motion can be solved and used to
eliminate $h_{\alpha\beta}$ from the action,
thereby obtaining the
``Nambu form.''  This Nambu form still has reparametrization symmetry, which
can be used to impose the Virasoro conditions as gauge
conditions that supplement the equations of motion given previously.

The T duality group O(6,22) relates Kaluza--Klein excitations
of the compactified string to winding-mode excitations.  From the point
of view of the conventional 2D world-sheet field theory, the KK
excitations can be understood perturbatively (in the $\alpha'$
expansion), whereas the winding-mode excitations are nonperturbative
solitons.  If the characteristic size of the compact dimensions is
called $R$, these statements are reflected in the fact that the masses of
Kaluza--Klein excitations are proportional to $1/R$, whereas those of
winding-mode excitations are proportional to $R/\alpha'$.  Thus, the
latter become infinitely heavy in the weak coupling limit $\alpha'
\rightarrow 0$, a characteristic feature of solitons.  Given
these facts, it seems remarkable that the O(6,22)
symmetry is realized on the action!  Clearly, this  requires some explanation.
The internal components of the metric and the other moduli are of order
$R^2/\alpha'$ (times dimensionless numbers).  In the usual string action
only terms proportional to $R^2/\alpha'$ or $1/\alpha'$ appear.
However, the matrix $M^{ab}$ is constructed out of the internal metric
and its inverse.  Thus, it has pieces proportional to $(R^2/\alpha')^n$
for $n = - 1, 0, + 1$.  To understand the symmetries in question
perturbatively in $\alpha'$, we must
consider $R^2$ to be of order $\alpha'$.  This means that the $y^2$
terms in the O(6,22) symmetric action \twofive\ are strongly coupled and must
be
treated exactly.  Fortunately, since the $y$ dependence in eq. \twofive\
is quadratic, this is possible and
explains why a symmetry that relates perturbative
excitations to solitons can be realized in the action.

\medskip

\noindent{\bf 3. Bogomol'nyi Bound, Soliton Spectrum, and Five-Branes}

The effective field theory of massless bosonic fields for heterotic
string theory compactified on a Narain torus at a generic point
in the moduli space can be written in various classically equivalent forms.
One form that has manifest T duality and general coordinate invariance is
$$\eqalign{
S =\ &{1\over 32\pi}
\int d^4 x\sqrt{-g}\Big[R -{1\over 2(\lambda_2)^2}
g^{\mu\nu}\p_\mu\lambda \p_\nu\bar\lambda -\sum_{a,b=1}^{28}
{\lambda_2\over 4}
F^a_{\mu\nu} (LML)_{ab} F^{b\mu\nu}\cr
&+{\lambda_1\over 4} \sum_{a,b=1}^{28} F^a_{\mu\nu} L_{ab} \tF^{b\mu\nu}
+{1\over 8} g^{\mu\nu}Tr(\p_\mu M L \p_\nu M L)\Big].\cr
}
\eqn\ebbzero
$$
The overall multiplicative factor of $1/32\pi$ is irrelevant for
classical analysis, and was omitted in
ref.\RSCHSEN, but it provides a convenient normalization of the
action when discussing charge quantization, breaking of SL(2,R)
symmetry to SL(2,Z), and the Bogomol'nyi bound\RDYONO\RDYONT.
Although there are no massless charged fields in this theory, the full
string theory does contain massive charged states, as well as soliton
states carrying magnetic charges.
The electric and magnetic charges $q_{el}^a$ and $q_{mag}^a$ of a
state are defined by \foot{These definitions
differ from those of
ref.\RDYONO\RDYONT\ by a factor of two due to different normalization
conventions for the gauge fields.}
$$
2 q^a_{el}=\lim_{r\to\infty} r x^iF^a_{0i}, \quad
2 q^a_{mag}=\lim_{r\to\infty} r x^i \tilde F^a_{0i}.
\eqn\ebbone
$$
The Bogomol'nyi lower bound\RBOGOM\
on the mass squared of a state for a given value of
$(q_{el}^a$, $q_{mag}^a)$ is given by\RHARLIU\RDYONT
$$
m^2 \geq {\lambda_2^{(0)}\over 16}
\Big(q_{el}^a(LM^{(0)}L+L)_{ab}q_{el}^b
+q_{mag}^a (LM^{(0)}L+L)_{ab} q_{mag}^b\Big)\equiv (m_0)^2,
\eqn\ebbtwo
$$
where the superscript $^{(0)}$ denotes the asymptotic value of the
corresponding field.

In ref.\RDYONT\
the expression for $m_0$ in eq.\ebbtwo\ was shown to be SL(2,Z) invariant.
In order to rewrite it in a manifestly SL(2,Z) invariant form, let us
express $q_{el}^a$ and $q_{mag}^a$ in terms
of vectors $\alpha_0^a$ and
$\beta_0^a$\RDYONT
$$
q_{el}^a={1\over \lambda_2^{(0)}}M^{(0)}_{ab}
(\alpha_0^b+\lambda_1^{(0)}\beta_0^b), \quad
q_{mag}^a=L_{ab}\beta_0^b,
\eqn\ebbthree
$$
where both $\alpha_0^a$ and $\beta_0^a$ belong to a reference lattice $P_0$,
which is
even and self-dual with respect to the metric $L$.
Now eq.\ebbtwo\ may be rewritten as
$$
(m_0)^2={1\over 16} \pmatrix{\alpha_0^a & \beta_0^a\cr}
\cm^{(0)}
(M^{(0)}+L)_{ab} \pmatrix{\alpha_0^b\cr \beta_0^b \cr},
\eqn\ebogom
$$
where we define
$$
\cm={1\over\lambda_2}\pmatrix{1 & \lambda_1\cr \lambda_1 &
|\lambda|^2\cr} \quad {\rm and} \quad \cl=\pmatrix{0 & 1\cr -1 & 0\cr}.
\eqn\ebogomaa
$$
$\cm$ and $\cl$ play the same role for SL(2,Z) that $M$ and $L$ do for
O(6,22;Z). Eq.\ebogom\ is manifestly invariant under
SL(2,Z) transformations $\omega$ and under O(6,22;Z)
transformations $\Omega$:\foot{The SL(2,Z)
and O(6,22;Z) transformation laws of the
vectors $\alpha_0^a$ and $\beta_0^a$ can be read off from eqs.\ebbone,
\ebbthree, and the known transformation laws\RDUALITY\RSCHSEN\
of the fields $\lambda$, $M$ and
$F^a_{\mu\nu}$ under these transformations.}
$$\eqalign{
M\to & \Omega^T M \Omega, \quad \alpha_0^a\to
(\Omega^{-1})_{ab}\alpha_0^b, \quad \beta_0^a\to
(\Omega^{-1})_{ab}\beta_0^b\cr
\cm\to & \omega^T \cm \omega, \quad
\pmatrix{\alpha_0^a\cr \beta_0^a\cr}\rightarrow \omega^{-1}
\pmatrix{\alpha_0^a\cr \beta_0^a\cr}.\cr
}
\eqn\ebbfour
$$

Eq.\ebogom\ suggests that it is natural to combine the vectors
$\alpha_0^a$ and $\beta_0^a$ into a single 56-dimensional vector
$\xi=\pmatrix{\alpha_0^a\cr \beta_0^a\cr}$
which
now belongs to a 56-dimensional lattice $\Gamma$. The new lattice
$\Gamma$ is self-dual not only with respect to the metric $L$, but also
with respect to the metric $\hcl=\cl\otimes L$.
The latter condition says that, for any two
vectors $\xi=(\alpha_0^a, \beta_0^a)$ and $\xi'=(\alpha_0^{\prime a},
\beta_0^{\prime a})$ belonging to the lattice $\Gamma$,
\foot{In fact, the terms are separately integers, since $P_0$ is even and
self-dual. This reflects the fact that there are states in the spectrum
without magnetic charge.}
$$ \xi^T \hcl \xi' =
\alpha_0^a L_{ab} \beta_0^{\prime b} - \alpha_0^{\prime a}L_{ab}
\beta_0^b ={\rm
integer}.
\eqn\equant
$$
In our normalization this is just the
DSZW quantization condition for the magnetic charge.

The statement that the spectrum of electric and magnetic charges in the
theory remains invariant under SL(2,Z) transformations\RDYONO\
can now be
translated to the statement that the lattice $\Gamma$ is invariant under
SL(2,Z) transformations.
This follows from eq.\ebbfour\ and the fact that both $\vec\alpha_0$
and $\vec\beta_0$ belong to the lattice $P_0$.
Similarly,
T duality invariance of the spectrum is the statement that $\Gamma$ is
invariant under O(6,22;Z) transformations.
This follows from the invariance of the lattice $P_0$ under such
transformations.

To summarize, we have expressed the mass squared of
supersymmetric states in the theory in a form that it is manifestly
invariant under the SL(2,Z) and
O(6,22;Z) transformations. Furthermore, these two transformations appear
on an equal footing. In this formalism,
both the SL(2,Z) and O(6,22;Z) invariances of the allowed spectrum
of charges correspond to the invariance of the lattice $\Gamma$ under
the corresponding transformations.
However, in string theory, there is a fundamental difference between
these two transformations.
An O(6,22;Z) transformation relates Kaluza--Klein modes to string
winding modes, and hence transforms perturbative string excitations to
perturbative string excitations, whereas an SL(2,Z) transformation
transforms perturbative string excitations to monopole (or dyon)
solutions in string theory. Thus, the spectrum of
perturbative string excitations has O(6,22;Z) symmetry, but not SL(2,Z)
symmetry. This can be seen explicitly by noting that the perturbative string
spectrum contains charge vectors $\vec\xi$ of the form
$\pmatrix{\alpha_0^a \cr 0\cr}$. States with $\beta_0^a \neq 0$
are solitons, and their masses diverge in the weak coupling limit.

Assuming that SL(2,Z) is a genuine symmetry of string theory,
it is reasonable to ask if there is some dual formulation of the theory
for which the spectrum of perturbative excitations has SL(2,Z) invariance,
and O(6,22;Z) symmetry of the spectrum becomes manifest only after
including the soliton solutions of this dual theory.
Let us now look for such a possibility among $p$-brane theories
in ten dimensions.
When a ten-dimensional $p$-brane theory is compactified on a torus to four
dimensions, the spectrum
includes the
usual Kaluza--Klein modes, which can be identified with the Kaluza--Klein
modes in string theory.
But there are also excitations that correspond to the
$p$-brane wrapped around the six-torus, which are required to be
the SL(2,Z) transforms of the Kaluza--Klein modes, just
as the string winding modes are
O(6,22;Z) transforms of Kaluza--Klein modes.

We shall now show that if such a scenario holds, $p$ must be five.
{}From eq.\ebogom, taking
$\lambda_1^{(0)}=0$ (i.e., vanishing $\theta$ angle),
the ratio of the masses of a
purely electrically charged particle to a purely magnetically charged
particle is given by
$$
1/\lambda_2^{(0)}=\lim_{r\to\infty}
e^{\Phi^{(10)}}\Big({\det G^{(10)}_{Smn}}\Big)^{-1/2}
\propto R^{-6},
\eqn\eccone
$$
where $\Phi^{(10)}$ is the ten-dimensional dilaton field,
$G^{(10)}_{Smn}$ denotes the internal components of the ten-dimensional
string
metric, and $R\propto \sqrt{G^{(10)}_{Smn}}$ denotes the linear scale of the
internal manifold measured in this metric.
Let us now consider the dependence of this mass ratio on $R$ for
a fixed value of $\Phi^{(10)}$.

For a $p$-brane theory compactified on a six-torus,
the masses of the Kaluza--Klein modes are proportional to
$1/R'$, whereas those of the $p$-brane winding modes, which are supposed
to be identified with the string theory monopoles, are proportional to
$R^{\prime p}$.
Here $R'$ denotes the radius of the internal manifold computed in the
$p$-brane metric, which can differ from the string metric by a
multiplicative factor involving the dilaton field. In order to study
the dependence of the mass ratio on $R$ for fixed value
of the dilaton field, we can take $R'$ to be proportional to $R$.
The mass ratio is then proportional to $R^{-p-1}$.
Comparison with the calculation based on string theory in eq.\eccone\ then
gives $p=5$.
This shows that if there exists a dual version of string theory for
which the perturbative spectrum is manifestly SL(2,Z) invariant, it must be a
theory of five-branes.

This result can be made more concrete by identifying the quantum
numbers $\alpha_0^m$ and $\beta_0^m$ ($1\le m\le 6$) with
the internal momenta and winding numbers of the five-brane wrapped
around a six-torus.
In this analysis all fields that arise from the
dimensional reduction of the 16 ten-dimensional gauge fields are set to zero,
and we only
consider states that do not carry any charge associated
with these gauge fields. In this case, the
indices $a,b$ in eq.\ebbzero\ can be taken to run from 1 to 12,
$\alpha^a_0$, $\beta^a_0$ can be regarded as 12-dimensional vectors, and
$M$ and $L$ can be taken to be 12$\times$12 matrices of the form
$$
M=\pmatrix{\hG^{-1} & \hG^{-1}\hB\cr -\hB\hG^{-1} & \hG - \hB \hG^{-1}
\hB\cr}, \quad L=\pmatrix{0 & I_6\cr I_6 & 0\cr},
\eqn\ecctwo
$$
where $\hG$ and $\hB$ are internal components of the metric and
antisymmetric tensor fields, respectively.
As was shown in ref.\RSCHSEN, the gauge field dependent part of the
action \ebbzero\ can be replaced by
$$\eqalign{-{1\over 128\pi}
\int d^4x \sqrt{-g}
\sum_{m,n=1}^6 \Big[ & F^{(m,\alpha)}_{\mu\nu}\hG_{mn}
(\cl^T\cm\cl)_{\alpha\beta}F^{(n,\beta)\mu\nu}
+ F_{\mu\nu}^{(m,\alpha)}\hB_{mn}\cl_{\alpha\beta}
\tF^{(n,\beta)\mu\nu}\Big].}
\eqn\eccthree
$$
The precise relation between the fields $F^{(m,\alpha)}_{\mu\nu}$ and
$F^a_{\mu\nu}$ can be found by using the manifestly
O(6,6)$\times$SL(2,R) form of the action given in ref.\RSCHSEN\ and
the equations of motion derived from that action.
This form of the action contains 24 field strengths
$F^{(a,\alpha)}_{\mu\nu}$ ($1\le a\le 12$, $1\le\alpha\le 2$), with
the identification
$$
F^{(a,1)}_{\mu\nu}=F^a_{\mu\nu}.
\eqn\eccfour
$$
The equations of motion relate $F^{(a,2)}$ to $F^{(a,1)}$. In
particular,
$$\eqalign{
F^{(m,2)}_{\mu\nu}=&\lambda_1 F^{(m,1)}_{\mu\nu}+\lambda_2
\hG^{mn}\tF^{(n+6,1)}_{\mu\nu} +\lambda_2
\hG^{mn}\hB_{np}\tF^{(p,1)}_{\mu\nu}\cr
=&\lambda_1 F^{m}_{\mu\nu}+\lambda_2
\hG^{mn}\tF^{n+6}_{\mu\nu} +\lambda_2
\hG^{mn}\hB_{np}\tF^{p}_{\mu\nu},
\quad \quad 1\le m\le 6.\cr
}
\eqn\eccsix
$$

Let us now consider adding source terms of the form
$$
{1\over 4}
\int d^4 x \sqrt{-g} (A_\mu^{(m,1)} J^\mu_m + A_\mu^{(m,2)} \tJ^\mu_m)
\eqn\eccseven
$$
to the action \eccthree.
For asymptotically Minkowskian metric $g_{\mu\nu}$ the gauge
field equations of motion derived from the combined action \eccthree,
\eccseven\ give rise to the following form of Gauss's law after we use
eqs.\ebbone, \ebbthree, \eccfour\ and \eccsix
$$\eqalign{
\int d^3 x\sqrt{-g} J^0_m = \ & {1\over 2} \lim_{r\to \infty} r x^i
({|\lambda|^2\over\lambda_2}
\hG_{mn} F^{(n,1)}_{0i} -{\lambda_1\over \lambda_2}
\hG_{mn} F^{(n,2)}_{0i} +\hB_{mn} \tF^{(n,2)}_{0i})=\alpha^m_0,\cr
\int d^3 x\sqrt{-g} \tJ^0_m = \ & {1\over 2}\lim_{r\to \infty} r x^i
({1\over \lambda_2}\hG_{mn} F^{(n,2)}_{0i} -{\lambda_1\over \lambda_2}
\hG_{mn}F^{(n,1)}_{0i} - \hB_{mn} \tF^{(n,1)}_{0i})=\beta_0^m.\cr
}
\eqn\eccnine
$$
This shows that the quantum numbers $\alpha_0^m$
and $\beta_0^m$ are the total charges coupled to the gauge fields
$A_\mu^{(m,1)}$ and $A_\mu^{(m,2)}$, respectively.
Since these gauge fields couple naturally to the
five-brane\RSCHSEN, the contribution to these charges from a given
configuration of the five-brane can be calculated.
To do this, let us introduce the
world-volume metric $\gamma_{rs}$ ($0\le r,s\le 5$) and ten-dimensional
fields $G^{(10)}_{FMN}$,
$\ca^{(10)}_{M_1\ldots M_6}$ ($0\le M,N,M_i\le 9$)
that couple naturally to the five-brane\RDUFF\RSTROM,
and write the five-brane $\sigma$-model action in terms of these
background fields:
$$
\int d^6 \xi [ {1\over 2} \sqrt{-\gamma} \gamma^{rs} G^{(10)}_{FMN}
\p_r Z^M \p_s Z^N  - 2\sqrt{-\gamma}
+ {1\over 6!}
\ca_{M_1\ldots M_6} \epsilon^{r_1\ldots r_6}
\p_{r_1} Z^{M_1} \ldots \p_{r_6} Z^{M_6}].
\eqn\eccten
$$
In writing this equation, the coupling of the
ten-dimensional dilaton field $\Phi^{(10)}$ to the five-brane has been
ignored, but this will not affect the analysis.
Let us now consider backgrounds characterized by non-zero values of
$$\eqalign{
& G^{(10)}_{Fmn}, \quad G^{(10)}_{F\mu\nu}, \quad \ca^{(10)}_{m_1\ldots
m_6}=\lambda_1 \epsilon_{m_1\ldots m_6},
\quad 1\le m,n \le 6, \quad \mu, \nu =0, 7,8,9\cr
}
\eqn\ecceleven
$$
with all other components of all the fields set to zero.
Denoting the internal coordinates by $Y^m$ and the space-time
coordinates by $X^\mu$, the action can be written as
$$\eqalign{
\int d^6\xi [{1\over 2} \sqrt{-\gamma}  & \gamma^{rs}
(G^{(10)}_{Fmn}\p_r Y^m \p_s Y^n
+G^{(10)}_{F\mu\nu}\p_r X^\mu \p_s X^\nu) + {\lambda_1\over 6!}
\epsilon_{m_1\ldots m_6} \epsilon^{r_1\ldots r_6} \p_{r_1} Y^{m_1}
\ldots \p_{r_6} Y^{m_6}]. \cr}
\eqn\eddtwo
$$
Taking the background fields to be independent of the internal
coordinates $Y^m$, this theory has the following two
conserved world-volume current densities
corresponding to internal momentum and winding-number densities of
the five-brane
$$\eqalign{
j^r_m= & (\sqrt{-\gamma}\gamma^{rs} G^{(10)}_{Fmn}\p_s Y^n +{\lambda_1\over
5!}
\epsilon^{r r_2\ldots r_6}\epsilon_{m m_2\ldots m_6} \p_{r_2}
Y^{m_2} \ldots \p_{r_6} Y^{m_6}), \cr
\tilde j^r_m = & {1\over 5!}
\epsilon^{r r_2\ldots r_6}\epsilon_{m m_2\ldots m_6} \p_{r_2}
Y^{m_2} \ldots \p_{r_6} Y^{m_6}.\cr
}
\eqn\eddtwoa
$$
Let us now introduce background fields $G^{(10)}_{Fm\mu}$ and
$\ca^{(10)}_{\mu m_2\ldots m_6}$ and write down the extra terms that appear
in the world-volume action to linear order in these fields.
Using the identifications\RSCHSEN
$$
G^{(10)}_{Fm\mu} = G^{(10)}_{Fmn} A^{(n,1)}_\mu, \quad
\ca^{(10)}_{\mu m_2\ldots m_6} = \epsilon_{m m_2\ldots m_6}
(-A^{(m,2)}_\mu +\lambda_1 A^{(m,1)}_\mu),
\eqn\eddthree
$$
the extra terms in the world-volume action take the form
$$
\int d^6\xi \Big(A^{(m,1)}_\mu j_m^r \p_r X^\mu
- A^{(m,2)}_\mu \tj_m^r \p_r X^\mu\Big).
\eqn\eddfive
$$
Let us now work in the static gauge $X^0=\xi^0$. Comparing
eqs.\eccseven\ and \eddfive, and using eqs.\eccnine, gives
$$
\alpha_0^m=4\int j_m^0 d^5\xi, \quad \beta_0^m =-4
\int \tj_m^0 d^5\xi,
\quad 1\le m\le 6.
\eqn\eddsix
$$
This shows that $\alpha_0^m$ and $\beta_0^m$ are proportional to the total
internal momenta and winding numbers of the five-brane, respectively.
In other words, the spectrum of perturbative five-brane excitations, which
contains both Kaluza--Klein states and five-brane winding states, is
characterized by states for which the first six components of the
twelve-dimensional vectors $\alpha^a$ and $\beta^a$ are non-zero. Since
this is an SL(2,Z) invariant set, the spectrum of allowed
charges for perturbative five-brane excitations is SL(2,Z) invariant.

This analysis does not prove definitively that the mass
spectrum of perturbative five-brane states is SL(2,Z) invariant. A
complete answer to this question requires a better
understanding of the five-brane mass spectrum. However,
the five-brane theory is characterized by the same space-time
supersymmetry algebra that is responsible for the Bogomol'nyi bound
\ebogom. (The presence of central charges
corresponding to five-brane winding numbers
in the supersymmetry algebra was established in
ref.\RTOWN.) Hence the masses of the perturbative five-brane
states are expected to satisfy the same lower bound as given in eq.\ebogom.
It remains to be proved that there are five-brane states that saturate this
bound.

We conclude this section with the observation that the perturbative five-brane
spectrum does not contain string winding modes; these must appear as
soliton solutions in the five-brane theory. Hence T duality is
not a symmetry of the perturbative five-brane spectrum.

\medskip

\noindent{\bf 4. SL(2,Z) Transformed World-Sheet Theories}

The preceding section described a spectrum of electric and
magnetic charge excitations in terms of 56-component vectors
$(\vec\alpha_0,\vec\beta_0)$
that is consistent with the S and T dualities of the
toroidally compactified theory. It was obtained by
saturating the Bogomol'nyi bound and is consistent with the most
general DSZW quantization requirements.  In perturbation theory (i.e.,
the expansion in Newton's constant), all of the electrically charged
states $(\vec\alpha_0,0)$ are present in the
spectrum, whereas none of the
magnetically charged states $(\vec\beta_0
\not= 0)$ appear.  All $\vec\beta_0 \not=
0$ states must arise nonperturbatively as solitons.  The world-sheet
theory of section 2 accounts for all the electrically charged states.
As was explained there, some of these are perturbative and some are
solitons from the world-sheet (first quantization) viewpoint.  However,
they are all perturbative from the space-time (second quantization)
viewpoint.  Since the S duality group SL(2,Z) relates electrically
charged states to magnetically charged states, it relates perturbative
states and nonperturbative states of the space-time theory, just as the
T duality group O(6,22;Z) did for the world-sheet theory.
Specifically, the SL(2,Z) group element $\left({a\ b\atop c\ d}\right)$ maps
states
with charges $(\vec\alpha_0,0)$ to ones with charges $(a\vec
\alpha_0, c \vec\alpha_0)$.
It is possible to find group elements for any pair of relatively prime
integers $a$ and $c$.

Now let us consider applying the SL(2,Z) transformation directly to the
world-sheet theory.  This transformation  acts nontrivially on the background
gauge fields
$A_\mu^a (x)$, as well as the antisymmetric tensor $B_{\mu\nu} (x)$ and the
dilaton $\Phi(x)$.  In the form that the theory has been written, these
are complicated nonlocal transformations.  However, this doesn't really
matter; the formulas are not required.  In terms of the
transformed fields $\tilde{A}(x), \tilde{B}(x)$, and $\tilde{\Phi}(x)$,
the transformed world-sheet
theory is isomorphic to the original one expressed in terms of $A,B,$
and $\Phi$. Therefore, this gives an S transformed dual formulation of
the world-sheet theory, for which the excitation spectrum has charge
vectors of the form $(a\vec\alpha_0, c\vec\alpha_0)$,
as measured by the original
gauge fields $A_\mu^a (x)$.\foot{For
long
strings, explicit soliton solutions representing these dual strings, and
some properties of these solutions were discussed in ref.\RSLTZ.}
Of course, from the viewpoint of the
transformed potentials $\tilde{A}_\mu^a$ these are electrically charged
states as before.  In terms of the transformed world-sheet theory, all
states with charges that are not of the form $(a\vec\alpha_0,
c\vec\alpha_0)$ must
arise as solitons of the associated space-time theory.
Therefore there are an
infinite number of equivalent dual starting points for defining the
theory, which can be labeled by pairs of relatively prime integers
$(a,c)$.  This generalizes the proposal of Montonen and Olive\ROLIVE\
(in another context) that there
should be two dual formulations, which could be called ``electric
strings'' $(a = 1, c = 0)$ and ``magnetic strings'' $(a = 0, c = 1)$.
Note that distinct dual formulations are labeled by
pairs $(a,c)$ rather than by SL(2,Z) elements $\left({a\ b\atop
c\ d}\right)$.  The reason for this is that the element $\left({1\ 1\atop
0\ 1}\right)$, which corresponds to a quantized shift of the axion
field, is a symmetry of the world-sheet theory.  Only group elements
with $c \not= 0$ act nontrivially on the electric string.

If one considers five-branes, on the other hand, one finds that the
perturbative excitations can be described as having six nontrivial
electric charges (the first six components of $\vec\alpha_0$) and six
nontrivial magnetic charges (the first six components of $\vec\beta_0$).
This charge
spectrum does have SL(2,Z) symmetry, but it is not O(6,22;Z) invariant.
Dual formulations of the five-brane world-volume theory can be reached by
acting with T
dualities in the manner described above for S dualities and
the string world-sheet theory.

\medskip

\noindent{\bf 5. Discussion}

By assuming that the heterotic string compactified on a
torus to four dimensions has S duality, this Letter has shown that
an attractive picture, satisfying a number of consistency tests, emerges.
It seems likely that this
symmetry is sufficiently robust that it is applicable even for more realistic
compactification schemes.
A deeper understanding of S duality should be helpful for
understanding  crucial features of realistic models, such as the origin of
supersymmetry breaking together with the absence of a cosmological
constant. It might even provide helpful clues for constructing a better
string field theory. Of course, we are still very far from such a level of
understanding.

There are some more modest, but still challenging, problems that may
be appropriate to study first: One is to generalize our analysis to
nongeneric values of the moduli for which there is unbroken nonabelian
gauge symmetry.  Another (possibly related) one is to explore whether
it is possible to construct an effective four-dimensional space-time theory
with S duality symmetry when charged states are included.
As has been explained,
perturbative string excitations should be related to
nonperturbative solitons by the S duality symmetry.  In Ref.\RSCHSEN\
a space-time action with SL(2,Z) symmetry was constructed, but this was
done only for the low-energy field theory without charged particle
excitations.  If any of them are added, then the magnetically charged states
that they transform into would need to be added, too.
One reason for  thinking that this might be possible is the example
of the world-sheet theory. As we
have shown, the world-sheet
theory can be recast so as to incorporate the T duality symmetry that relates
perturbative Kaluza--Klein excitations to winding-mode solitons.

Some related issues have been discussed in recent papers by
Kallosh and Ortin, Bin\'etruy, and
Duff and Khuri\RRECENT. We gratefully acknowledge useful
discussions with P. Bin\'etruy, M. Duff, and A. Strominger.

\refout
\end